\documentclass[runningheads]{llncs}

\usepackage[mobile]{eccv}

\usepackage{eccvabbrv}

\usepackage{graphicx}
\usepackage{booktabs}

\usepackage[accsupp]{axessibility} 

\usepackage[pagebackref,breaklinks,colorlinks]{hyperref}

\usepackage{orcidlink}

\usepackage{blindtext}

\usepackage{amsmath,amsfonts,bm}
\usepackage{nicefrac}

\def\density{\rho}
\def\stiffs{\kappa_s}
\def\stiffb{\kappa_b}
\def\Image{\mathbf{I}}

\def\ImageTC{\mathbf{I}_{1:T,1:C}}

\def\Verts{\mathbf{V}}
\def\Faces{\mathbf{F}}

\def\texture{\rmT}
\newcommand{\Loss}[1]{\mathcal{L}_{\textrm{#1}}}
\def\render{\mathcal{R}}
\def\ours{PhysAvatar{}}

\def\Tableref#1{Table~\ref{#1}}

\def\Figref#1{Fig.~\ref{#1}}

\def\secref#1{section~\ref{#1}}

\def\eqref#1{eq.~\ref{#1}}

\def\1{\bm{1}}

\def\rvc{{\mathbf{c}}}

\def\rvf{{\mathbf{f}}}

\def\rvn{{\mathbf{n}}}

\def\rvp{{\mathbf{p}}}
\def\rvq{{\mathbf{q}}}

\def\rvs{{\mathbf{s}}}

\def\rvv{{\mathbf{v}}}

\def\rmT{{\mathbf{T}}}

\DeclareMathAlphabet{\mathsfit}{\encodingdefault}{\sfdefault}{m}{sl}
\SetMathAlphabet{\mathsfit}{bold}{\encodingdefault}{\sfdefault}{bx}{n}

\newcommand{\R}{\mathbb{R}}

\usepackage{booktabs} 
\usepackage{paralist}
\usepackage{makecell}
\usepackage{multirow}
\usepackage{multicol}
\usepackage{flushend}
\usepackage{afterpage}

\usepackage{array}
\newcolumntype{L}[1]{>{\raggedright\let\newline\\\arraybackslash\hspace{0pt}}m{#1}}
\newcolumntype{C}[1]{>{\centering\let\newline\\\arraybackslash\hspace{0pt}}m{#1}}
\newcolumntype{R}[1]{>{\raggedleft\let\newline\\\arraybackslash\hspace{0pt}}m{#1}}

\usepackage[font=small,skip=2pt]{caption}

\usepackage[ruled]{algorithm2e} 

\SetAlFnt{\small}
\SetAlCapFnt{\small}
\SetAlCapNameFnt{\small}
\SetAlCapHSkip{0pt}

\usepackage{bbold}

\usepackage{xpunctuate}

\begin{document}

\title{PhysAvatar: Learning the Physics of Dressed 3D Avatars from Visual Observations}

\titlerunning{PhysAvatar}

\author{Yang Zheng$^{1}\thanks{Equal Contribution}$ \and
Qingqing Zhao$^{1\star}$ \and
Guandao Yang$^{1}$ \and
Wang Yifan$^{1}$ \and
Donglai Xiang$^{2}$ \and
Florian Dubost$^{3}$ \and
Dmitry Lagun$^{3}$ \and
Thabo Beeler$^{3}$ \and
Federico Tombari$^{3,4}$ \and
Leonidas Guibas$^{1}$ \and Gordon Wetzstein$^{1}$ 
}

\authorrunning{Yang Zheng et al.}

\institute{Stanford University \and
Carnegie Mellon University \and
Google \and
Technical University of Munich
}

\maketitle

\begin{abstract}
Modeling and rendering photorealistic avatars is of crucial importance in many applications. Existing methods that build a 3D avatar from visual observations, however, struggle to reconstruct clothed humans. We introduce PhysAvatar, a novel framework that combines inverse rendering with inverse physics to automatically estimate the shape and appearance of a human from multi-view video data along with the physical parameters of the fabric of their clothes. For this purpose, we adopt a mesh-aligned 4D Gaussian technique for spatio-temporal mesh tracking as well as a physically based inverse renderer to estimate the intrinsic material properties. PhysAvatar integrates a physics simulator to estimate the physical parameters of the garments using gradient-based optimization in a principled manner. These novel capabilities enable PhysAvatar to create high-quality novel-view renderings of avatars dressed in loose-fitting clothes under motions and lighting conditions not seen in the training data. This marks a significant advancement towards modeling photorealistic digital humans using physically based inverse rendering with physics in the loop. Our project website is at: \url{https://qingqing-zhao.github.io/PhysAvatar}.
\keywords{neural rendering \and physics \and dynamic modeling  \and 3D avatar}
\begin{figure}[!h]
    \centering    
    \includegraphics[width=\linewidth]{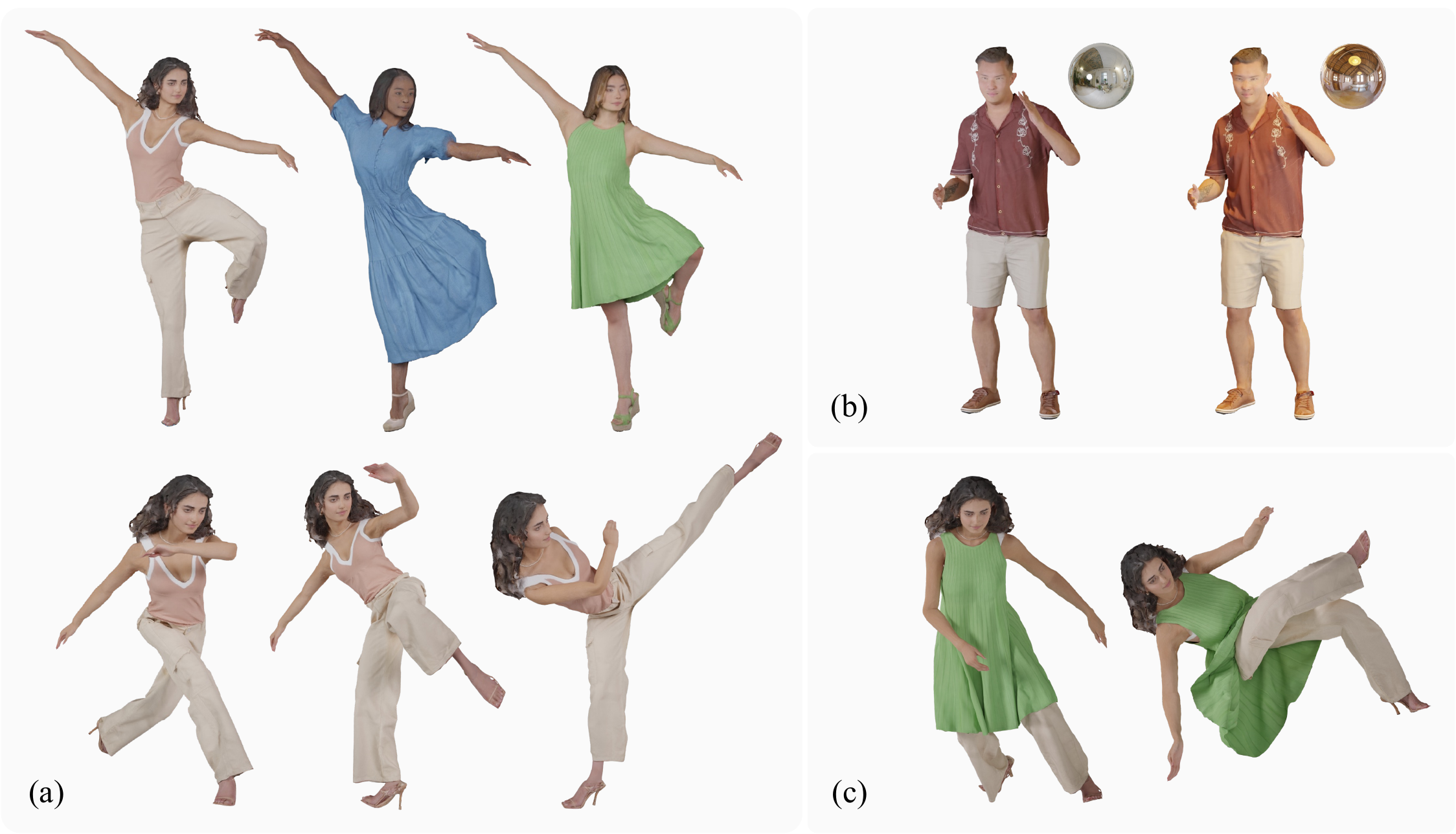}
    \caption{PhysAvatar is a novel framework that captures the physics of dressed 3D avatars from visual observations, enabling a wide spectrum of applications, such as (a) animation, (b) relighting, and (c) redressing, with high-fidelity rendering results.}
    \vspace{-15pt}
    \label{fig:teaser}
\end{figure}

\end{abstract}

\section{Introduction}
\label{sec:intro}

Digital avatars are a vital component in numerous applications, ranging from virtual reality and gaming to telepresence and e-commerce~\cite{genay2021being,sun2022human,korban2022survey}.
A realistic 3D avatar of a person can be readily obtained from visual observations, such as multi-view image~\cite{wang2023rodin} and video~\cite{bashirov2024morf} data. 
The task of rendering animated 3D avatars from novel viewpoints or when performing unseen motions, however, presents considerable challenges, particularly when the avatar wears loose-fitting garments. 
Accurately rendering the dynamics of garments in conditions that are not observed in the training data necessitates a holistic approach that not only models the shape and appearance of the person but also the physical behavior of their clothes, including friction and collision. 

Learning 3D scene representations from visual data is the core problem addressed by inverse rendering---a field that has recently shown remarkable progress in estimating the geometry and appearance of static and dynamic scenes from multi-view images and videos~\cite{Tewari2020State,tewari2022advances}. In the context of reconstructing 3D avatars, the ability to explicitly control the motions of the avatar in post processing becomes imperative.
Most existing methods for reconstructing animatable avatars drive the animation through an underlying skeleton using linear blending skinning (LBS) \cite{loper2015smpl}, which adequately models the dynamics of humans dressed in tight-fitting clothes~\cite{su2021nerf,bergman2022generative,noguchi2022unsupervised,Dong_2023_ICCV,xu2023efficient,abdal2023gaussian}. With this approach, garments are treated as a rigidly attached part of the body adhering to piece-wise linear transformations, which results in motions that appear rigid and unconvincing in many cases~\cite{Noguchi_2021_ICCV,zielonka2023drivable,xu2023gaussian}. Several successful attempts to introduce non-rigid deformations through pose-dependent geometry adjustments, such as wrinkles, have recently been made~\cite{liu2019neural, liu2021neural,wang2022arah,li2023posevocab,chen2023uv}, although these methods are prone to overfitting to the body poses and motion sequences observed during training. A key problem is that most existing 3D avatar reconstruction approaches neglect to model the dynamics of loose garments in a physically accurate manner, leading to unrealistic fabric behavior and issues like self-penetration. To our knowledge, only the work of Xiang \etal~\cite{xiang2022dressing} includes a physics-based approach to inverse rendering of digital humans, but their approach requires a tedious manual search process to find the parameters that model the dynamic behavior of their cloth fabrics with reasonable accuracy. 

Here, we introduce \ours, a novel approach to 3D avatar reconstruction from multi-view video data. \ours{} combines inverse rendering with ``inverse physics'' in a principled manner to not only estimate the shape and appearance of the avatar but also physical parameters, including density and stiffness, of the fabrics modeling their (loose) clothes. Our approach includes several parts:
Given the reconstructed 3D mesh of the garment in one frame, we first leverage mesh-aligned 4D Gaussians~\cite{wu20234d} to track surface points of the garment across all frames of the input video sequence. This process establishes dense correspondences on the 3D garment. These data are used as supervision in an inverse physics step, where the physical parameters of the fabric are optimized using a finite-difference approach~\cite{yee1997finite,leveque2007finite}.
Finally, we employ a physically based inverse renderer~\cite{nimier2019mitsuba} to jointly estimate ambient lighting and the surface material. By leveraging the refined geometry from our simulation step, the inverse renderer can effectively factor in pose-dependent effects, such as shadows caused by self-occlusion, resulting in accurate appearance reconstruction that enables the avatar to be rendered in novel lighting conditions. 

\ours{} offers a comprehensive solution for reconstructing clothed human avatars from multi-view video data to perform novel view and motion synthesis with state-of-the-art realism.
The key contributions of our work include:
\begin{compactenum}
\item The introduction of a new inverse rendering paradigm for avatars created from real-world captures that incorporates the physics of loose garments in a principled manner;
\item A pipeline that includes accurate and efficient mesh reconstruction and tracking using 4D Gaussians; automatic optimization of the garments' physical material properties; and accurate appearance estimation using physically based inverse rendering.
\end{compactenum}
\ours{} demonstrates a novel path to learning physical scene properties from visual observations by incorporating inverse rendering with physics constraints. Code will be available upon publication.

\section{Related Work}
\label{sec:related}

\subsection{Scene Reconstruction from Visual Observations}
Reconstructing 3D scenes from visual observations is grounded in classic approaches, such as Structure from Motion (SfM)~\cite{snavely2006photo, schonberger2016structure,newcombe2011kinectfusion}. 
Recently, this field has witnessed a paradigm shift towards inverse rendering techniques that estimate the shape and appearance of a scene from visual input~\cite{Tewari2020State,tewari2022advances}. Among the many representations developed in this area are neural (radiance) fields~\cite{mildenhall2021nerf,sitzmann2019srns}, neural volumes, surfaces, and signed distance fields~\cite{park2019deepsdf,mescheder2019occupancy,sitzmann2019deepvoxels,lombardi2019neural,wang2021neus,yariv2021volume} as well as differentiable rasterization techniques for meshes~\cite{Kato2017Neural,liu2019soft} and point clouds~\cite{yifan2019differentiable,lassner2021pulsar,keselman2022approximate,kerbl20233d}.
Recent inverse rendering approaches are capable of modeling dynamic scenes~\cite{Pumarola2020D-NeRF,Tretschk2020Non-Rigid,park2021hypernerf,fridovich2023k} and enable sophisticated post-capture deformations~\cite{yuan2022nerf,Jambon2023NeRFshop}.

\subsection{Animatable Avatars}
One central requirement for generating full-body digital humans is to enable animations driven by skeleton-level control~\cite{komatsu1988human,magnenat1988joint}. To this end, early works on avatar reconstruction primarily relied on skinning techniques, which successfully model the dynamics of bodies with minimal or tight clothing~\cite{anguelov2005scape,loper2015smpl}. Clothing is an integral part of everyday human appearance, captured by the saying ``clothes make the (wo)man''. Reconstructing their dynamic behavior from visual data, however, remains a challenge. Several different types of representations have been explored for clothed avatars, including meshes~\cite{ma2020learning} with dynamic textures \cite{bagautdinov2021driving,habermann2021real,zhao2022human}, neural surface~\cite{saito2021scanimate,chen2022gdna,tiwari2021neural} and radiance fields~\cite{peng2021neural,peng2021animatable,kwon2021neural,su2021nerf,chen2023uv,geng2023learning,kwon2023neural,kwon2023deliffas,chen2024anidress,zheng2022structured,feng2022capturing}, point sets~\cite{ma2021power,ma2021scale,ma2022neural}, and 3D Gaussians~\cite{hu2023gaussianavatar,li2023animatable,pang2023ash,zielonka2023drivable}. Many of these works condition the deformation of clothing on the body pose and some predict future dynamics based on a small number of frames~\cite{habermann2021real,habermann2023hdhumans,pang2023ash}.

We now discuss recent research most closely related to ours.
TAVA~\cite{li2022tava} learns a deformable neural radiance field of humans or animals using a differentiable skinning method like SNARF~\cite{chen2021snarf}.
ARAH~\cite{ARAH:2022:ECCV} models 3D avatars using SDF representations and proposes an SDF-based volume rendering method to reconstruct the avatar in canonical space. 
GS-Avatar~\cite{hu2023gaussianavatar} achieves real-time animation by leveraging a learned 3D Gaussian predictor. 
Although these methods show reasonable effectiveness on humans wearing tight garments, they struggle with loose garments because their LBS-based deformation module cannot accommodate the complex dynamics of loose clothing.
To address this limitation, Xiang \etal~\cite{xiang2022dressing} incorporate a cloth simulator based on the eXtended Position Based Dynamics (XPBD) formulation~\cite{xpbd} into a deep appearance model to reconstruct realistic dynamics and appearance. 
However, they require tedious manual adjustments of physical parameters to produce reasonable garment motion, and the code is not publicly available.
Concurrent with our work,
AniDress~\cite{chen2024anidress}, while using physics-based simulation, only uses the simulation to produce a garment rigging model. 
This can cause inaccuracy in cloth dynamics modeling and unrealistic motion artifacts.  
In contrast, we estimate the physical parameters through a principled gradient-based inverse physics approach and accurately recover the shape and (relightable) appearance using a physically based inverse renderer, achieving high-fidelity results on novel motions.

\subsection{Physics-Based Simulation}

Physics-based simulation of cloth \cite{baraff2023large,tang2013gpu,liu2017quasi,zeller2005cloth,vassilev2001fast,peng2023pgn,yu2019simulcap, wang2021gpu} has been extensively studied and is particularly useful for modeling complicated dynamic effects in interaction with human bodies, such as large deformations, wrinkling, and contact. In this work, we adopt the Codimensional Incremental Potential Contact (C-IPC) solver~\cite{li2021codimensional} for its robustness in handling complicated body--cloth collision by the log-barrier penalty function \cite{li2020incremental}. We refer readers to dedicated surveys \cite{jiang2008survey,li2023review} for a comprehensive review, and focus on works that solve the inverse parameter estimation problem. One line of work uses specialized devices to simultaneously measure forces and deformation in a controlled setting \cite{wang2011data,miguel2012data,larionov2022estimating,zhang2024estimating}. Our work falls into the category that estimates physical parameters directly from imagery of garments worn on human subjects \cite{bhat2003estimating,stoll2010video,guo2021inverse}. For this purpose, some differentiable simulators have been developed and applied to the optimization problem~\cite{li2023diffavatar}, but only for specific types of solvers, such as Projective Dynamics \cite{li2022diffcloth} and XPBD \cite{stuyck2023diffxpbd}. 
The heavy runtime cost of high-quality physics-based simulation motivates research in \textit{neural cloth simulation}. Early works create datasets of garments simulated on human bodies \cite{bertiche2020cloth3d,gundogdu2019garnet}, and then train the network to predict garment deformation in a supervised manner \cite{santesteban2019learning,chentanez2020cloth,vidaurre2020fully,pan2022predicting}. Recently, more attention has been paid to the self-supervised formulation that applies the elasticity energy defined in traditional simulation to predict garments, thereby training the network to act like a simulation solver, including both the (quasi-) static \cite{bertiche2021deepsd,bertiche2021pbns,santesteban2022snug,lee2023clothcombo,de2023drapenet} and the dynamic \cite{bertiche2022neural,grigorev2023hood} cases. Most of these works focus on making forward predictions that resemble traditional simulators without solving inverse parameter estimation on real-world garment data, except CaPhy \cite{su2023caphy}. CaPhy estimates the Saint Venant-Kirchhoff (StVK) elasticity parameters of clothing from 4D scans but only predicts pose-dependent clothing deformation in a quasi-static manner. In addition, all the works mentioned above treat simulation as a separate problem without modeling the photorealistic appearance. By comparison, our work builds avatars with physically based appearance and dynamics that are optimized for faithfulness to the real human capture in a holistic manner.

\section{Method}
\label{sec:method}
\begin{figure}[t!]
    \centering
    \includegraphics[width=\linewidth]{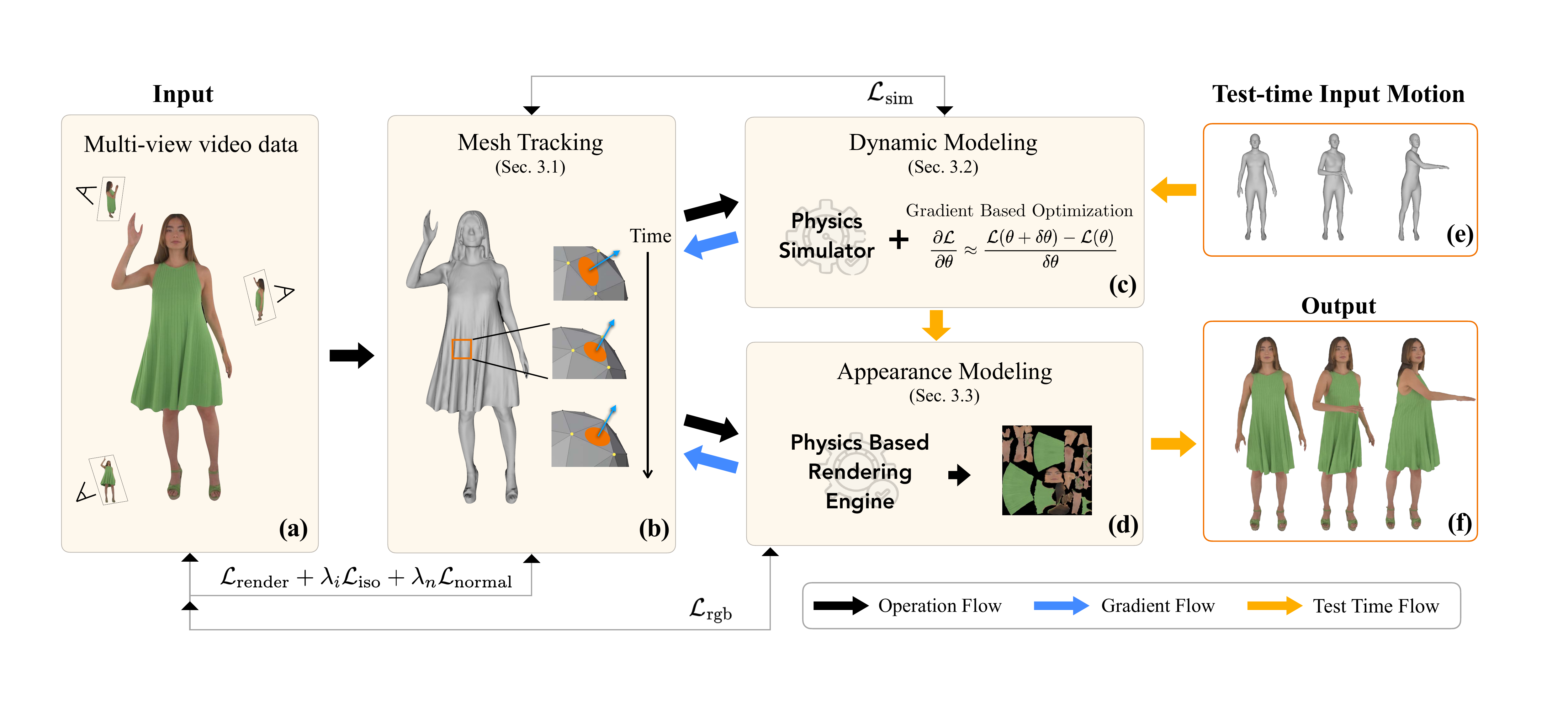}
    \caption{\textbf{Method Overview:} (a) \ours{} takes multi-view videos and an initial mesh as input.  We first perform (b) dynamic mesh tracking (Sec.~\ref{sec:method-mesh}). The tacked mesh sequences are then used for (c) garment physics estimation with a physics simulator combined with gradient-based optimization (Sec.~\ref{sec:method-phy}); (d) and appearance estimation through physics-based differentiable rendering (Sec.~\ref{sec:method-render}). At test time, (e) given a sequence of body poses (f), we simulate garment dynamics with the learned physics parameters and employ physics-based rendering to produce the final images.}
    \label{fig:method}
\end{figure}
Given multi-view videos as input, our goal is to reconstruct a 3D avatar that can be realistically animated and rendered under novel motions and novel views. The core pipeline (Fig.~\ref{fig:method}) consists of three primary components:
\begin{inparaenum}
\item Mesh tracking (Sec.~\ref{sec:method-mesh})---given multi-view videos and a reconstructed mesh at the initial time step as input, we track the deformation of the geometry, which provides the geometry reference for the ensuing physics simulation; 
\item Physics-based dynamic modeling (Sec.~\ref{sec:method-phy})---with the tracked meshes as reference, we estimate the garments' physical properties using a physics simulator; the simulator, together with the estimated physical parameters are used to generate novel motion;
\item Physics-based appearance refinement (Sec.~\ref{sec:method-render})---given the tracked geometry, we further enhance the appearance estimation through a physics-based differentiable renderer, which considers self-occlusion and thus eliminate artifacts such as baked-in shadows.
\end{inparaenum}
In the following, we delve into the details of each component.

\subsection{Mesh Tracking}\label{sec:method-mesh}
Given multiview videos of \(T\) frames and \(C\) views,  $\left\lbrace\ImageTC\right\rbrace$, and an initial mesh \((\Verts_1, \mathbf{F})\), with vertices $\Verts_1$ and faces $\mathbf{F}$, that can be obtained via any existing static scene reconstruction methods, our goal is to accurately track the mesh deformation through the video sequence \(\Verts_{1:T}\), which will be used as reference in the subsequent physics simulation.
However accurately tracking mesh deformation from visual observation is very challenging.
Inspired by the robust and real-time tracking performance from dynamic 3D Gaussians~\cite{luiten2023dynamic}, we use 3D Gaussians as a surrogate representation for mesh tracking.

We adapt 3D Gaussians to align with the reconstructed mesh surface.
Following the definition of dynamic 3D Gaussians, at frame \(t\), each Gaussian is parameterized with position \(\rvp_{t}\in\R^{3}\), rotation quaternion \(\rvq_{t}\in\R^{4}\), color \(\rvc_{t}\in\R^{3}\), scale \(\rvs_{t}\in\R^{3}\) and opacity \(o_{t}\in\R_{+}\).
To couple the Gaussians with the mesh, we attach one Gaussian at the barycenter of each face $\mathbf{f}\in\Faces$, and determine the Gaussians' quaternions from the attached triangle.
Specifically, the rotation of each Gaussian is computed from the rotation of its local frame, whose $x$- and \(z\)-axis are defined using the longest triangle edge at \(t=1\) and the face normal respectively.
Furthermore, we set the last scaling factor, which corresponds to the scaling in face normal direction, to be a small constant value, ensuring the Gaussian lies on the mesh surface.
The Gaussians can then be rendered in a differentiable manner~\cite{kerbl20233d} and we optimize their parameters by minimizing the loss~\cite{luiten2023dynamic} between the rendered images and the reference images:
\begin{equation}
    \Loss{render}=\lambda\left\|\Image_{i,t}-\hat{\Image}_{i,t} \right\|_1+ \left(1-\lambda\right)\cdot \text{SSIM}\left(\Image_{i,t}, \hat{\Image}_{i,t}\right),
    \label{loss:render}
\end{equation}
where $\hat{\Image}_{i,t}$ denotes the image rendered from the perspective of the $i$-th camera at time $t$.
Note that since the Gaussians' position and rotation are bound to the mesh, the optimization of the mesh vertices will implicitly optimize the parameters of the Gaussians. 
At the first time step, we optimize color, scale, and opacity.
In the following frames \(1<t\leq T\), we fix the opacity and scale, and optimize Gaussian colors \(\rvc_{t}\) and mesh vertices \(\Verts_t\).

\begin{figure}[!t]
    \centering
    \includegraphics[width=0.9\linewidth]{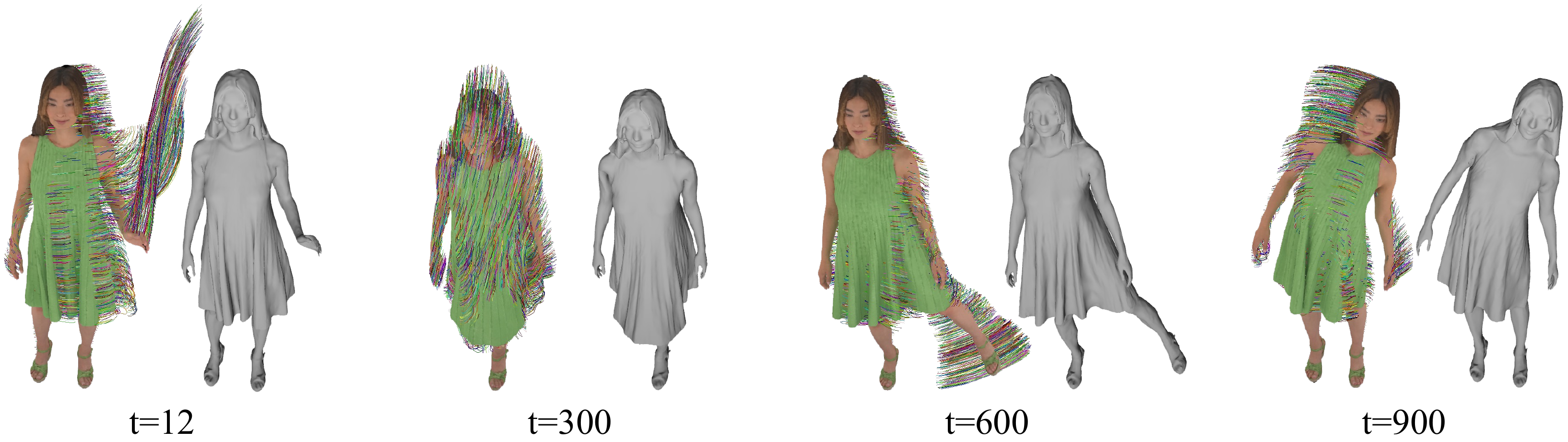}
    \caption{Our method can robustly track a dynamic mesh from input images, providing accurate long-term correspondences. Here we show the rendered images overlaid with the Gaussian trajectories from the previous 12 frames and the optimized meshes.} 
    \vspace{-15pt}
    \label{fig:tracking}
\end{figure}
In addition to the photometric loss, we apply the following regularization terms to preserve the local geometric features:
1) an isometry loss \(\Loss{iso}\)~\cite{luiten2023dynamic} that preserves edge length through the training sequence, 2) a normal loss \(\Loss{normal}\) that encourages smooth mesh surfaces,  enhancing mesh deformation accuracy and stability.
Denoting the vertex and the triangle of the mesh at frame \(t\) as \(\rvv_t\) and \(\rvf_{t}\), the normal of a face as \(\rvn\left( \cdot \right)\), and the 1-ring neighbors \(N\left(\cdot\right)\), the losses are formally defined as:
\begin{align}
    \Loss{iso}&=\frac{1}{\left|\Verts\right|}\sum_{\rvv\in\Verts}\sum_{\rvv'\in N\left( \rvv_{1} \right)}\frac{\left(\left\Vert\rvv_{1} - \rvv'_{1}\right\Vert_2 - \left\Vert\rvv_{t} - \rvv'_{t}\right\Vert_2\right)}{\left|N\left( \rvv \right)\right|},\label{loss:iso}\\
    \Loss{normal}&=\frac{1}{\left|\Faces\right|}\sum_{\rvf_{t}\in\Faces}\sum_{\rvf_{t}'\in N\left(\rvf\right)}\left(1 - \rvn\left(\rvf_t\right) \cdot \rvn\left(\rvf'_t\right)\right). \label{loss:normal}
\end{align}

The full loss used for mesh tracking is defined as:
\begin{equation}
    \Loss{mesh}=\Loss{render}+\lambda_i\Loss{iso} + \lambda_n\Loss{normal},
\end{equation}
where we set $\lambda_i=10$ and $\lambda_n=0.1$ during our training. With the introduced losses, our method can reconstruct a smooth mesh sequence with accurate correspondences, as shown in Fig.~\ref{fig:tracking}.

\subsection{Physics Based Dynamic Modeling}
\label{sec:method-phy}
Given the tracked meshes, $(\Verts_{1:T}, \Faces)$, we estimate the physical properties of the garments such that the simulation of their dynamics matches the reference tracked meshes.
Once learned, these parameters can be used to simulate the garment under novel motions and novel views.

\vspace{3mm}\noindent\textbf{Garment Material Model and Simulation.}
We model the elastodynamics of the garment by equipping each simulation mesh with the discrete shell hinge bending energy~\cite{grinspun2003discrete,tamstorf2013discrete} and isotropic StVK elastic
potential~\cite{chen2018physical,clyde2017modeling}.
Specifically, we estimate density $\density$, membrane stiffness $\stiffs$, and bending stiffness $\stiffb$ of the garment given the visual observations.
For our generic use cases, it is reasonable to assume homogeneous physical properties for each garment piece, i.e., the density, membrane stiffness, and bending stiffness can be parameterized using scalar values, denoted as  $\density$,  $\stiffs$, and $\stiffb$.
In choosing the simulators, we consider differentiability and quality. Unfortunately, the development of differentiable simulators suitable for our application remains an open research question. Current implementations, such as those documented in \cite{li2022diffcloth}, cannot be directly applied due to limitations in modeling complex body colliders.
Therefore, we propose to use the state-of-the-art robust and
stable cloth simulator C-IPC \cite{li2021codimensional} that guarantees no interpenetration with frictional contact. While C-IPC is differentiable, it lacks an implementation of accurate analytical gradients; hence, we use finite difference methods to estimate the gradients for parameter updates.
Specifically, we employ an open-source implementation of C-IPC~\cite{li2021codimensional} as our cloth simulator, which is a state-of-the-art simulator achieving high-fidelity garment simulation with stability and convergence (see \cite{li2021codimensional} for a comprehensive review of the base algorithm and implementation).

\vspace{3mm}\noindent\textbf{Garment Material Estimation.}
From the reconstructed mesh, we segment out the garment vertices to obtain ground truth garment mesh sequences $(\widehat{\Verts}_{1:T}^g, \widehat{\Faces}_{\phantom{f}}^g)$.
The cloth simulator, $f(\cdot)$, takes as input the current garment state---position and velocity ($\Verts^g_{t}$, $\dot{\Verts}^g_{t}$)---the garment's physical properties $(\density, \stiffs, \stiffb)$, 
the body collider $\{\Verts_{t}^C,\mathbf{F}^C\}$, and garment boundary vertices $\Verts_{t+1}^b$, and step size \(\Delta t\).
The boundary vertices of the garment are points of the garment that are in permanent contact with the human body, such as the shoulder region for draped garments. We employ the SMPL-X body model\cite{loper2015smpl} as the body collider $\{\Verts_{t+1}^C,\mathbf{F}^C\}$, extracting SMPL-X body shapes and poses from the video data during the pre-processing (see Sec.~\ref{sec:exp}).
The garment and the boundary vertices are manually annotated once on the initial mesh \(\Verts_{1}\) for the entire simulation. 
The simulation function predicting the garment shape in the next frame can be written as below: 
\begin{equation}
    \Verts^g_{t+1} = f(\Verts^g_{t}, \dot{\Verts}^g_{t}, \Verts_{t+1}^b, \mathbf{V}^C_{t+1}, \rho, \stiffs, \stiffb, \Delta t).
    \label{eq:simulator}
\end{equation}
We find the optimal physical parameters by minimizing the difference between the simulated and the reconstructed garment meshes. Formally, the optimization objective can be expressed as:
\begin{gather}
    \begin{aligned}
        \min_{\left\{\rho,\stiffs, \stiffb \right\}} & \Loss{sim}(\rho, \stiffs, \stiffb) = \sum_{t=0}^{T} \|\Verts^g_{t+1} - \widehat{\Verts}_{t+1}^{g}\|_2^2
        \label{eq:garment_opt}
    \end{aligned},
\end{gather}
where \(\Verts^g_{1} = \widehat{\mathbf{V}}^{g}_1\) and \(\Verts^g_{t+1}\) is computed by \cref{eq:simulator}.
We estimate gradients via a numerical approach using finite differences~\cite{leveque2007finite}.
This method is particularly feasible in our context due to the low-dimensional nature of the optimization problem, which involves only three scalar parameters.
We employ the forward difference formula for gradient estimation:
\begin{align}
    \frac{\partial \Loss{sim}}{\partial \rho} &\approx \left( \Loss{\text{sim}}(\rho+\delta_\rho, \stiffs, \stiffb) - \Loss{sim}(\rho, \stiffs, \stiffb) \right) / {\delta_\rho}.
    \label{eq:fd}
\end{align}
Where $\delta_\rho$ is a hyperparameter for forward difference gradient estimation. The gradients for other parameters are estimated using the same approach, and the parameters are subsequently updated using the Adam optimizer.

\vspace{3mm}\noindent\textbf{Animation.}
At test time, we have the initial mesh \{\(\Verts_1,\Faces_1\)\} and we are given a sequence of novel human body poses, e.g., provided by skeleton parameters such as SMPL~\cite{loper2015smpl}.
Using standard LBS, we obtain the boundary locations \(\Verts^b_{t+1}\). We assume the initial garment is at rest, i.e., $\dot{\Verts}_0 = 0$.
We use the same garment segmentation, boundary points, and SMPL body collider points as during training, and use \cref{eq:simulator} to simulate the deformation of the garment.
This is then combined with the body, driven by LBS, to create the holistic animation. 
\begin{figure}[!t]
    \centering
    \includegraphics[width=1\linewidth]{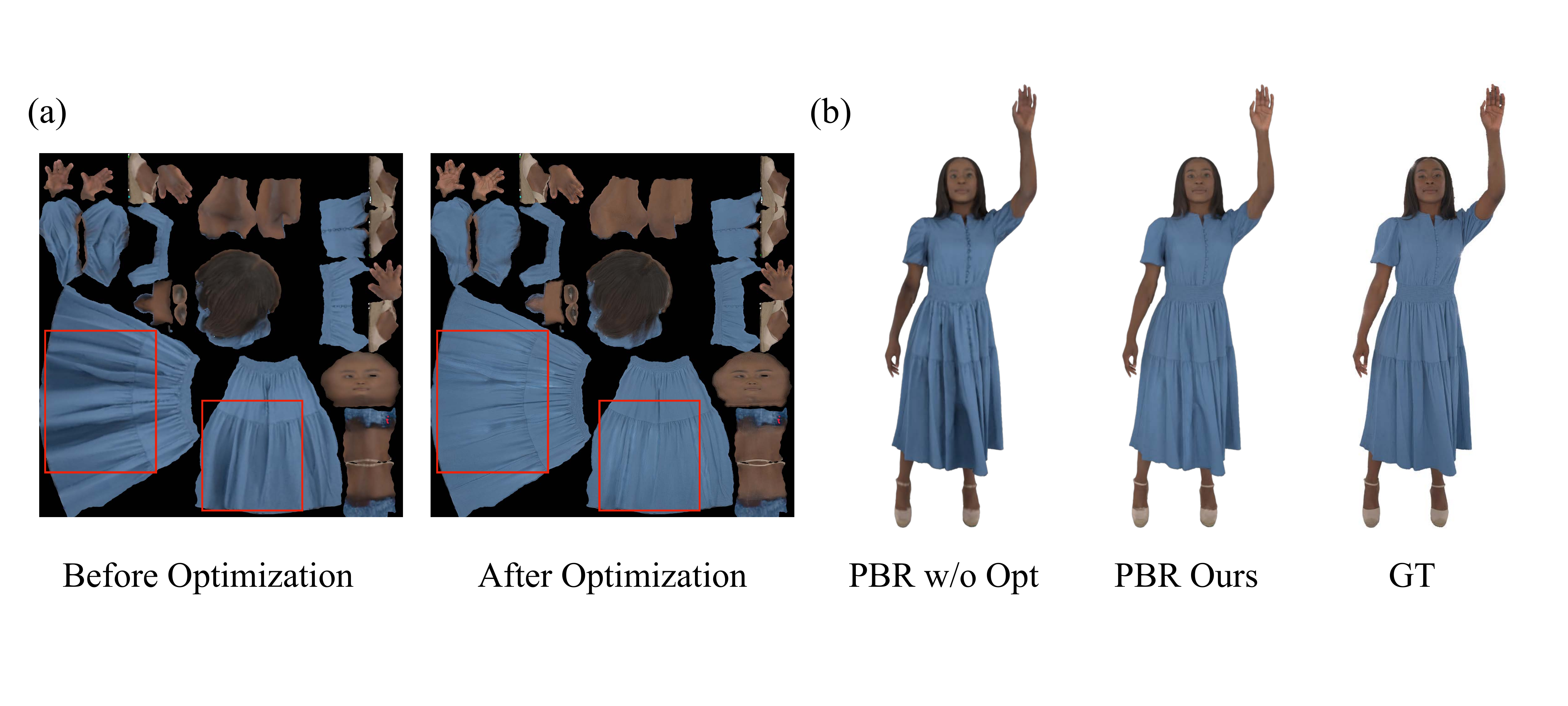}
    \vspace{-25pt}
    \caption{Ablation study on appearance estimation: (a) Initial texture map $\texture$ extracted from Gaussian splatting (left) has baked-in shadows highlighted in the red boxes; post-optimization (right), the baked-in shadows are substantially removed. (b) Rendering comparisons demonstrate that our method with the optimized texture map more closely aligns with the ground truth.}
    \vspace{-10pt}
    \label{fig:pbr}
\end{figure}
\subsection{Physics Based Appearance Modeling}\label{sec:method-render}
While tracking with dynamic Gaussians jointly estimates the garments appearance, it does not account for light--surface interactions, consequently the reconstructed appearance contains baked-in shadows, causing uncanny visual artifacts when rendered in novel views and motions (see Fig.~\ref{fig:pbr}).
In this step, given the mesh sequence \(\left(\Verts_{1:T}, \Faces\right)\) and the per-frame mesh colors \(\rvc_{1:T}\) reconstructed from \Cref{sec:method-mesh}, we utilize Mitsuba3~\cite{nimier2019mitsuba}---a physics-based differentiable Monte Carlo renderer---to enhance the appearance reconstruction.
For this purpose, we use a time-invariant diffuse texture map \(\texture\) to model the appearance. 
The value \(\texture\) is initialized from the average Gaussian colors from the tracking step (\Cref{sec:method-mesh}), $\nicefrac{\sum_{1<t\leq T}(\mathbf{c(x)}_t)}{T}$.
For the environment lighting, since the capture studio is typically equipped with uniform point light arrays, we can approximate the environment lighting as an ambient illumination, parameterized with a global \(L_{a}\in\R^{3}\).

We optimize for \(\texture\) and \(L_{a}\) using the photometric loss defined in~\cref{eq:pbr} for all training views and frames \(\left(i,t\right)\), where \(\render\) denotes the rendering function and \(c_{i,t}\) the camera pose:
\begin{align}   
 \Loss{rgb}=\min_{\left\{ \texture, L_a \right\}} \sum_{i,t}\left\|\Image_{i,t}-\render\left(\texture, L_a, \Verts_t, c_{i,t}\right)\right\|^2_2.
    \label{eq:pbr}
\end{align}
At test time, we can swap the ambient light with more complex illuminations, and produce realistic rendering in arbitrary views and body poses (see \cref{sec:application}).

\section{Experiments}
\label{sec:exp}

In this section, we explain our experimental setup and results. We recommend watching the supplementary video for better visualization.
\label{sec:results}

\begin{figure}[t!]
    \centering
    \includegraphics[width=\linewidth]{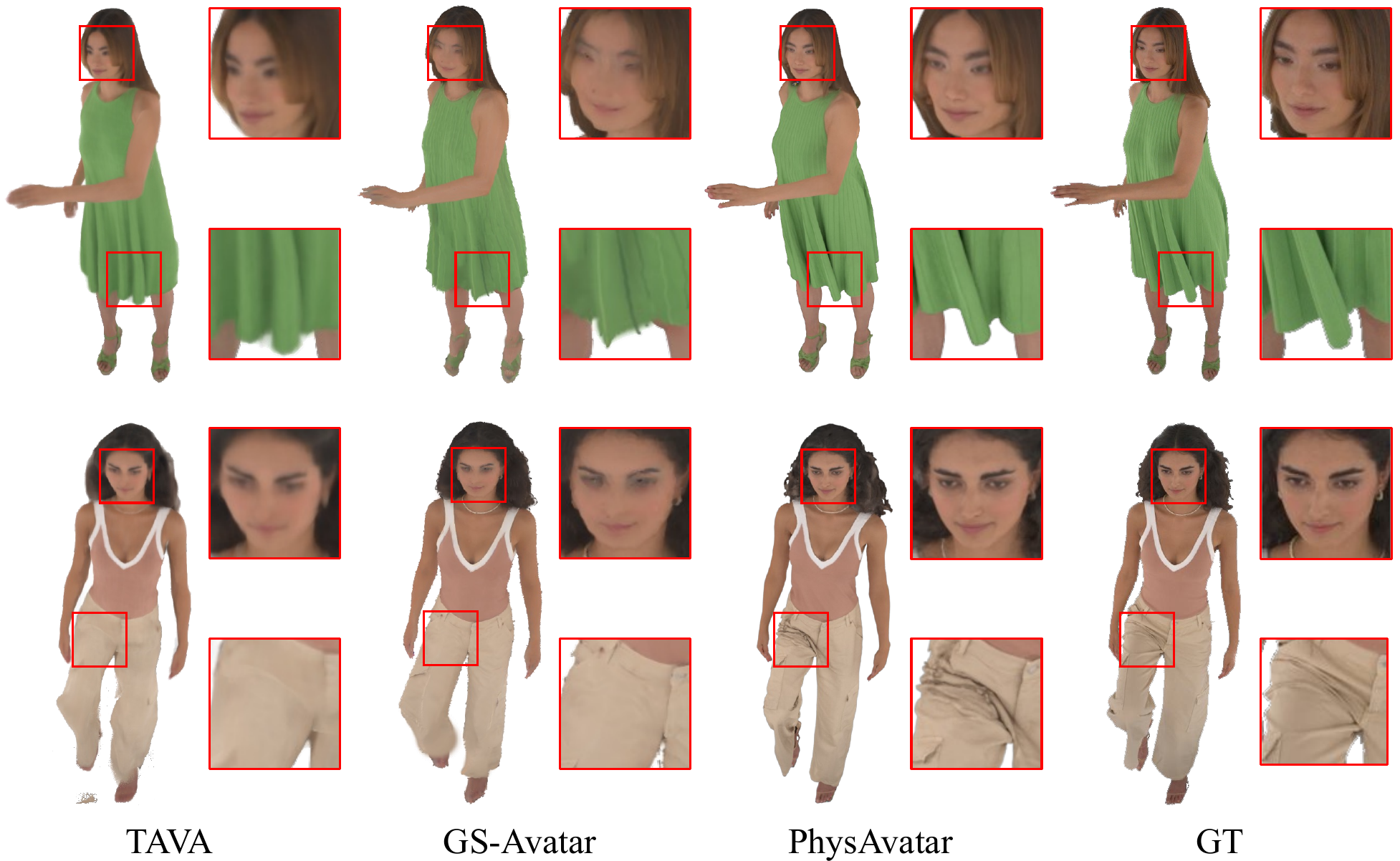}
    \caption{Qualitative results on test poses from the ActorHQ~\cite{jiang2023hifi4g} dataset. Our method PhysAvatar achieves state-of-the-art performance in terms of geometry detail and appearance modeling.  } 
    \vspace{-15pt}
    \label{fig:qualitative_baseline}
    
\end{figure}
\subsection{Experimental setup}
\textbf{Data Preparation.} We choose four characters from the ActorHQ~\cite{isik2023humanrf} dataset for our experiments, including two female characters with loose dresses, one female character wearing a tank top and loose-fitting pants, and one male character with a short-sleeved shirt and khaki shorts, as shown in Fig.~\ref{fig:teaser}. For each character, we use the ground truth mesh provided by the dataset and re-mesh it to obtain a simulation-ready initial mesh, i.e., one where the triangles have similar areas, to enhance the performance of garment simulation. 
We then segment the garment, define the boundary vertices of the garment for simulation purposes, and unwrap the whole mesh to get the UV map, all performed in Blender~\cite{blender}. Since we observe that the human pose estimations from the dataset are quite noisy, we fit the SMPL-X~\cite{SMPL-X:2019} body mesh using multi-view videos by minimizing the distance between the projected SMPL-X joints and 2D keypoint detection results from DWPose~\cite{mmpose2020, yang2023effective}.

\vspace{3mm}\noindent\textbf{Implementation Details.} For training, we utilize the entire set of 160 camera views from the ActorHQ~\cite{isik2023humanrf} dataset. Although the dataset includes approximately 2000 frames for each character, we selectively use 24 frames with large movements for estimating garment material parameters and 200 frames for optimizing the texture map, which leads to good results while conserving computational resources. 
For evaluation, we choose 200 frames for each character, with novel motions unseen during training. At test time, given a motion sequence defined by SMPL-X~\cite{SMPL-X:2019}, we use an LBS weight inpainting algorithm~\cite{abdrashitov2023robust} to obtain the skinning weights of the human body with respect to the SMPL-X skeleton. We animate the human body using LBS and simulate the garment, which is driven by the boundary vertices and utilizes the SMPL-X body mesh as the collider. More details are available at supplementary document. 

\vspace{3mm}\noindent\textbf{Baselines.} We benchmark our method and current open-sourced state-of-the-art methods modeling full-body avatars, including ARAH~\cite{ARAH:2022:ECCV} which leverages SDF-based volume rendering to learn the canonical representation of the avatar, TAVA~\cite{li2022tava} which learns the canonical radiance field~\cite{mildenhall2021nerf} and skinning weight of the character, and GS-Avatar~\cite{hu2023gaussianavatar} based on a pose-conditioned neural network which predicts the parameters of 3D Gaussians~\cite{kerbl20233d}. For all baselines, we use the official public codebase and extend them to our dense-view setting.
\begin{table*}[t!]
    \centering
    \caption{Quantitative comparison with baselines. \ours{} outperforms all baselines with respect to geometry accuracy. We achieve the best or competitive results in appearance metrics. See \Figref{fig:qualitative_baseline} and \Figref{fig:results_amass} for qualitative comparisons.} 
    \begin{tabular}{cccccc}
    \toprule
    \multirow{2}{*}{Method} & \multicolumn{2}{c}{Geometry} & \multicolumn{3}{c}{Appearance} \\ 
    & CD ($\times 10^{3}$) (\(\downarrow\)) & F-Score (\(\uparrow\)) & LPIPS (\(\downarrow\))  & PSNR (\(\uparrow\)) & SSIM (\(\uparrow\)) \\\midrule
    ARAH  \cite{ARAH:2022:ECCV}& 1.12 & 86.1 & 0.055  & 28.6 & \underline{0.957} \\
    TAVA  \cite{li2022tava}& \underline{0.66} & \underline{92.3} & 0.051 & 29.6 & \textbf{0.962}  \\
    GS-Avatar~\cite{hu2023gaussianavatar}& 0.91 & 89.4 & \underline{0.044} & \textbf{30.6} & \textbf{0.962} \\
    \ours{} (ours) & \textbf{0.55} & \textbf{92.9} & \textbf{0.035} & \underline{30.2} & \underline{0.957}  \\
    \bottomrule
    \end{tabular}
    \label{tab:quantitative}
\end{table*}
\subsection{Comparison}
\begin{figure}[t!]
    \centering
    \includegraphics[width=\linewidth]{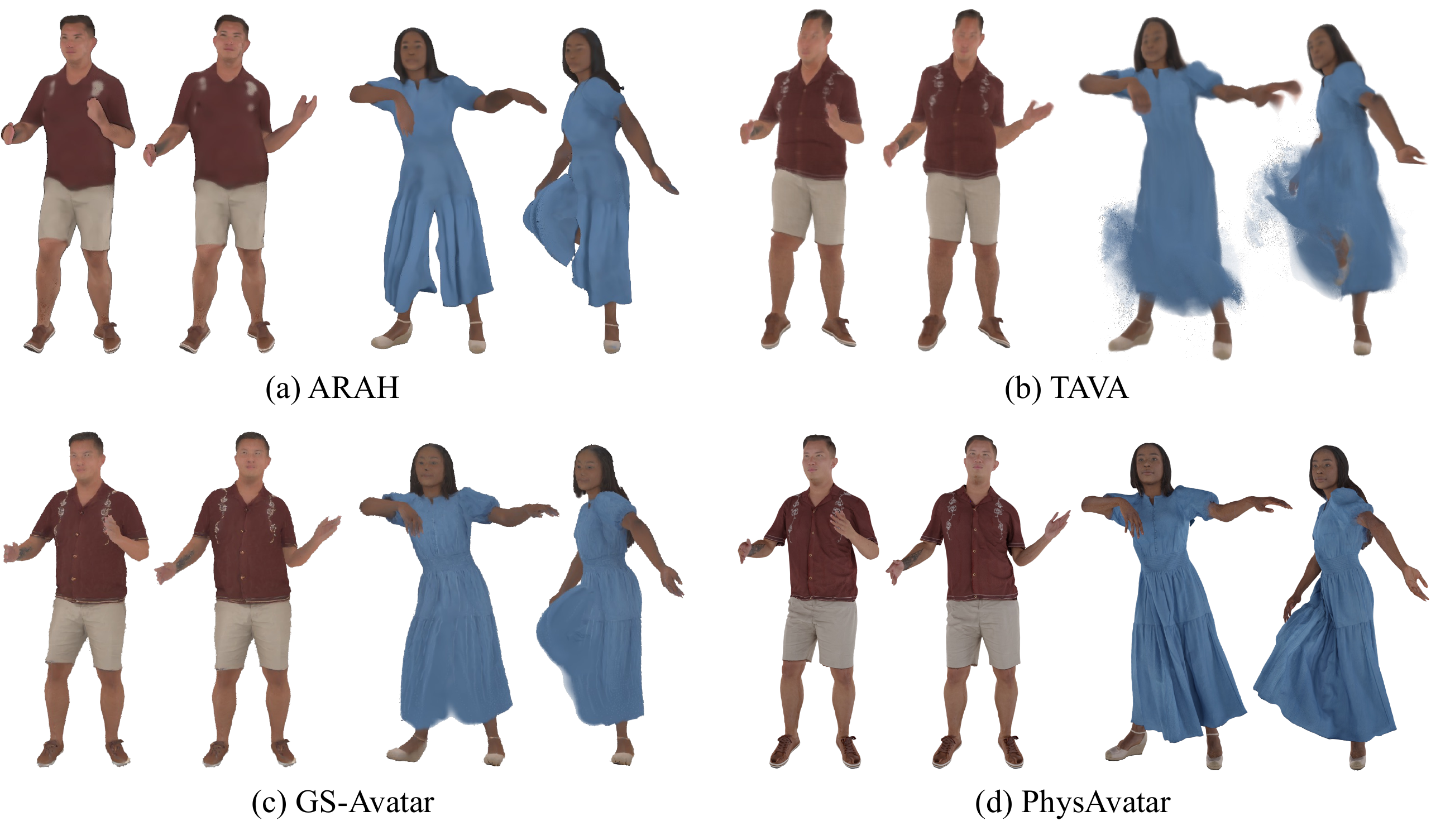}
    \vspace{-15pt}
    \caption{Here we show animation results of current state-of-the-art methods including ARAH~\cite{ARAH:2022:ECCV}, TAVA~\cite{li2022tava} and GS-Avatar~\cite{hu2023gaussianavatar}, and our results on test motions from AMASS~\cite{mahmood2019amass} dataset. Images are rendered from novel views. Please zoom in for details.}
    \vspace{-15pt}
    \label{fig:results_amass}
\end{figure}

We compare \ours{} against current state-of-the-art approaches in terms of geometry and appearance quality across test pose sequences, as shown in \Tableref{tab:quantitative}, \Figref{fig:qualitative_baseline}, and \Figref{fig:results_amass}.

\vspace{3mm}\noindent\textbf{Geometry Evaluation.} For geometry evaluation, we employ the mean Chamfer distance\cite{mescheder2019occupancy} and the mean F-Score\cite{tatarchenko2019single} at $\tau = 0.001$ as quantitative metrics. \ours{} outperforms all baseline according to these metrics, demonstrating a superior capability in accurately capturing the dynamic nature of garment geometry over time, as shown in \Tableref{tab:quantitative}. As shown in \Figref{fig:qualitative_baseline}, \ours{} generates more accurate garment deformation and better captures fine wrinkle details, thanks to the integration of a physics-based simulator. This contrasts with all baseline methods\cite{li2022tava,ARAH:2022:ECCV,hu2023gaussianavatar}, which do not explicitly model garment dynamics, but instead depend on learning quasi-static, pose-conditioned skinning weights or deformations. Additionally, we show results using motion data from the AMASS dataset\cite{mahmood2019amass}. \ours{} demonstrates its robustness by consistently synthesizing realistic garment deformation with fine details, while baseline methods sometimes fail to capture large deformations in loose garments (\Figref{fig:results_amass}).

\vspace{3mm}\noindent\textbf{Appearance Evaluation.}
For appearance evaluation, we utilize Peak Signal-to-Noise Ratio (PSNR), Structural Similarity Index Measure (SSIM), and the Learned Perceptual Image Patch Similarity (LPIPS)\cite{zhang2018unreasonable}. As detailed in \Tableref{tab:quantitative}, \ours{} surpasses all baselines in LPIPS, a metric shown to be better aligned with human perception\cite{zhang2018unreasonable}. While we achieve the second-best results in PSNR and SSIM, it is important to note that even minor misalignments in images can significantly impact these metrics, potentially not fully reflecting the quality of the renderings. As demonstrated in \Figref{fig:qualitative_baseline}, \ours{} can capture more high-frequency details in facial features and garment textures, thereby creating images that are visually richer and more detailed, compared to the baseline approaches.

\subsection{Ablation}

We perform ablation studies to assess the impact of each component in our pipeline on the overall performance. For this, we focus on two female characters from the ActorHQ dataset, specifically chosen for their loose garments.

\begin{figure}[!t]
    \centering
    \includegraphics[width=0.95\linewidth]{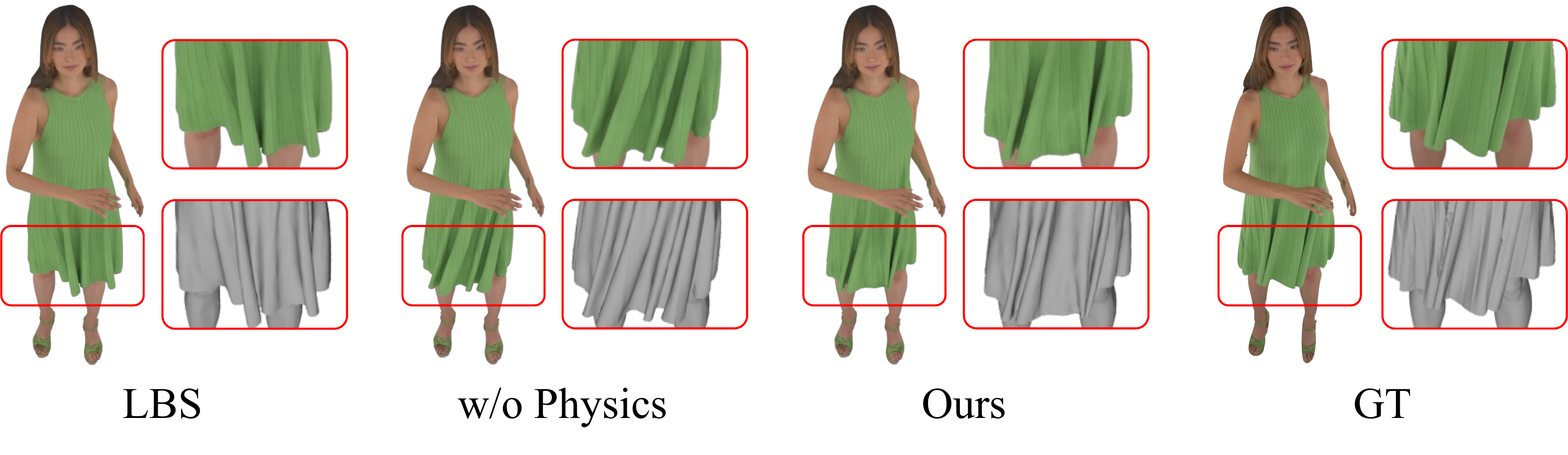}
    \vspace{-4pt}
    \caption{Ablation Study on garment physics estimation: LBS-based garment deformation (LBS) fails to generate realistic garment deformation. Physics-based simulation with random parameters (w/o physics) achieves realistic deformation but falls short of matching the ground-truth reference (GT). In contrast, \ours{} accurately captures garment dynamics, closely aligning with the reference.}
    \vspace{-0pt}
    \label{fig:ab_physics}
\end{figure}

\vspace{3mm}\noindent\textbf{Garment physics estimation and modeling.} We ablate our approach against two variants: one employing Linear Blend Skinning (LBS) for garment deformation and another utilizing a physics simulator for garment dynamics without optimized parameters, henceforth referred to as `w/o physics', which employs random garment parameters. As shown in \Tableref{tab:ablation}, both alternatives exhibit worse performance across all geometric metrics. Further, as shown in Fig.~\ref{fig:ab_physics}, LBS fails to produce plausible deformations for loose garments, such as dresses. While simulation with random parameters yields plausible garment dynamics, they do not accurately reflect real-world observations. In contrast, our method, with learned physics parameters, achieves deformations that closely align with the reference, demonstrating its effectiveness.

\vspace{3mm}\noindent\textbf{Appearance estimation and modeling.} We compare rendering outcomes with those obtained without appearance optimization through physics-based inverse rendering\cite{nimier2019mitsuba} (denoted as 'w/o appearance'), where the texture is directly extracted from the Gaussian framework, as detailed in \secref{sec:method-render}. \Figref{fig:pbr} illustrates that, since the Gaussian framework does not model complex light-surface interactions, shadows are inherently baked into the texture. However, post-optimization, we note a significant removal of these baked shadows. Quantitatively, we see improvements across all appearance metrics post-optimization, shown in \Tableref{tab:ablation}. 

\begin{table*}[t!]
    \centering
    \begin{tabular}{ccccccc}
    \toprule
    \multirow{2}{*}{Method} & \multicolumn{2}{C{0.35\linewidth}}{Geometry} & \multicolumn{3}{C{0.3\linewidth}}{Appearance} \\ 
    & CD ($\times 10^{-3}$) (\(\downarrow\)) & F-Score (\(\uparrow\)) & LPIPS (\(\downarrow\)) & PSNR (\(\uparrow\)) & SSIM (\(\uparrow\))  \\\midrule

    LBS             & 0.71          & 91.6           & 0.0354       & \underline{30.27}         & \textbf{0.951}         \\
    w/o physics     & \underline{0.61}          & \underline{92.2}           & \underline{0.0347}      & 30.26         & \textbf{0.951}          \\
    w/o appearance  & ---             & ---          & 0.0402         & 28.26         & \underline{0.947}     \\
    \ours{} (ours)    & \textbf{0.56} & \textbf{93.0}  &\textbf{0.0343} &\textbf{30.29} & \textbf{0.951}         \\

    \bottomrule
    \end{tabular}
    \vspace{5pt}
    \caption{Ablation Study: 
    \ours{} demonstrates improved geometric accuracy compared to both LBS-based animation (LBS) and simulations with random garment parameters (w/o physics), also shown in \Figref{fig:ab_physics}. \ours{} exhibits superior appearance metrics compared to the setting without physics-based appearance estimation (w/o appearance) (see \Figref{fig:pbr}).}
    \vspace{-5pt}
    \label{tab:ablation}
\end{table*}
\begin{figure}[t!]
    \centering
    \includegraphics[width=\linewidth]{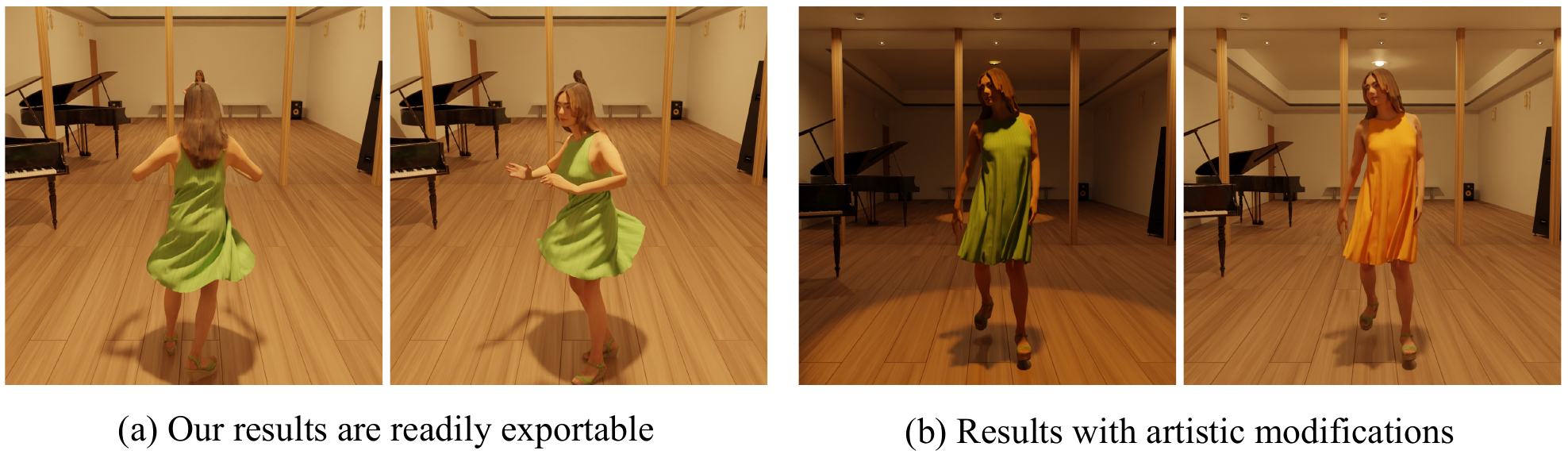}
    \caption{Our results can be exported into a traditional computer graphics pipeline, e.g., Blender~\cite{blender}, which enables the avatars performing novel motions to be rendered in unseen environments (a) as well as editing lighting and texture of the garments (b).}
    \label{fig:application}
    \vspace{-10pt}
\end{figure}
\subsection{Application}\label{sec:application} 
\ours{} can facilitate a wide range of applications, such as animation, relighting, and redressing, as demonstrated in Fig.~\ref{fig:teaser}. Our method can also benefit from other motion databases in addition to motions defined under the regime of SMPL-X~\cite{SMPL-X:2019} or SMPL~\cite{loper2015smpl}. Fig.~\ref{fig:teaser} shows our animation results on challenging motions from Mixamo~\cite{Mixamo}, where we use its auto-rigging algorithm to drive the human body, boundary vertices of the garment, and SMPL-X mesh as the collider. Current state-of-the-art approaches cannot handle such settings since they are mainly based on LBS weights from SMPL or SMPL-X. Moreover, while current methods relying on NeRF~\cite{mildenhall2021nerf} or 3D Gaussians~\cite{kerbl20233d} representation are usually incompatible with legacy graphics pipelines, our results are readily exportable and can be seamlessly integrated with computer graphics software tools (e.g., Blender~\cite{blender}), enabling post-processing artistic modifications (Fig.~\ref{fig:application}). 

\section{Limitations and Future Work}
Our method currently depends on manual garment segmentation and mesh UV unwrapping. We are posed to evolve our pipeline towards a more automated system by incorporating neural networks to streamline these data preprocessing tasks. Another limitation of our method is the absence of a perfect underlining human body for garment simulation. Our current use of the SMPL-X~\cite{SMPL-X:2019} model as the body collider introduces potential inaccuracies in collision detection and contact information due to mismatches between the SMPL-X mesh and the actual human body shape. Our method would benefit from using more advanced parametric models that promise a closer approximation to real human body shape. Furthermore, our pipeline leverages dense view captures to model avatars, an approach that, while effective, has its limitations. Adapting our method to a sparse-view setting would enhance the versatility of its applications in environments where capturing dense views is impractical. 
\section{Conclusion}
We introduce \ours, a comprehensive pipeline to model clothed 3D human avatars with state-of-the-art realism. By integrating a robust and efficient mesh tracking method, and a novel inverse rendering paradigm to capture the physics of loose garments and the appearance of the body, we believe our work represents a significant advancement in digital avatar modeling from visual observations, which also paves the way for innovative applications in entertainment, virtual reality, and digital fashion.
\section{Acknowledgement}
We would like to thank Jiayi Eris Zhang for the discussions. This material is based on work that is partially funded by an unrestricted gift from Google, Samsung, an SNF Postdoc Mobility fellowship, ARL grant W911NF-21-2-0104, and a Vannevar Bush Faculty Fellowship.





\setcounter{section}{0}
\renewcommand\thesection{\Alph{section}}
\newcommand{\suppsection}{\subsection}
\begin{flushleft}
\vspace{20pt}
\textbf{\Large Appendix}
\end{flushleft}
\makeatletter

\section{Implementation Details}
\subsection{SMPL-X Fitting}
\begin{figure}[t!]
    \centering
    \includegraphics[width=\linewidth]{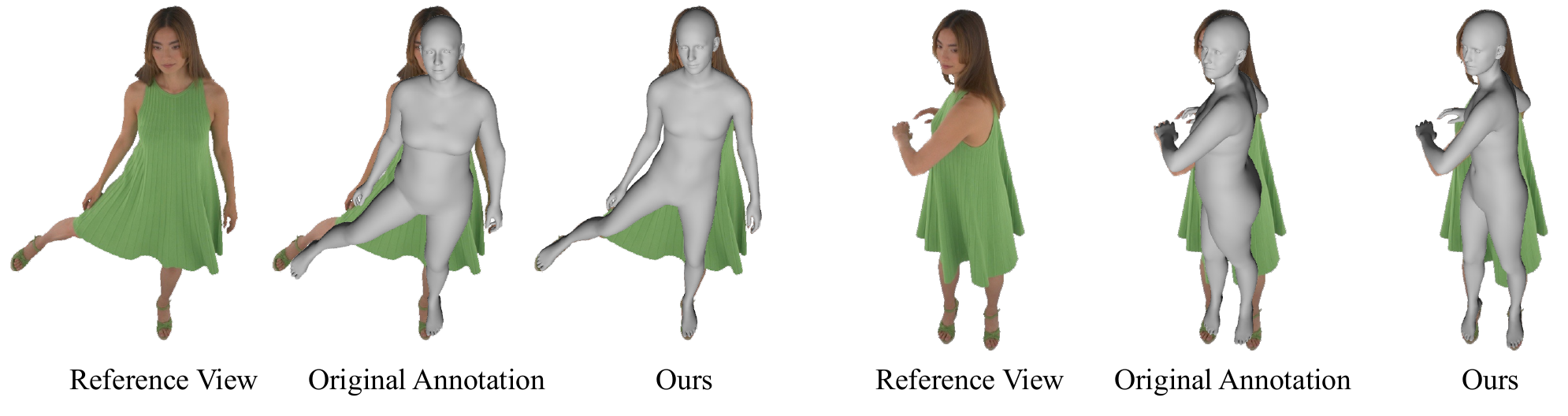}
    \vspace{-10pt}
    \caption{Our SMPL-X~\cite{SMPL-X:2019} fitting results show better accuracy compared to the original annotations from Actor-HQ~\cite{isik2023humanrf} dataset.}
    \label{fig:smplx}
\end{figure}

We observe that the human pose estimations provided by Actor-HQ~\cite{isik2023humanrf} are noisy, and failure cases exist in some sequences with challenging poses. Since our method relies on a body collider that provides accurate contact information for garment simulation, an SMPL-X mesh that aligns well with the human pose and shape is necessary. To this end, we fit the SMPL-X parameters from multi-view videos using an algorithm~\cite{zhang2021lightweight, zheng2021deepmulticap} simplified for our setting. Specifically, given image $\Image_{i,t}$ captured from camera $i$ at $t$ timestamp, we detect 2D human keypoints $\mathbf{x}_{i, t}$ using state-of-the-art human keypoints detector DWPose~\cite{mmpose2020, yang2023effective} and calculate the projection loss as follows:
\begin{equation}
    \mathcal{L}_{2d} = \left\| \mathbf{x}_{i, t} - \hat{\mathbf{x}}_{i, t} \right\|_2,
    \label{loss:projection}
\end{equation}
where $\hat{\mathbf{x}}_{i, t}$ is the projection of SMPL-X 3D joints under camera $i$. We also employ a regularization term as:
\begin{equation}
    \mathcal{L}_{reg} = \lambda_m\left\|m_h\right\|_2 + \lambda_{l}\left\|\theta_b\right\|_2 + \lambda_{f}\left\|\theta_f\right\|_2 + 
    \lambda_{\beta}\left\|\beta\right\|_2 + \lambda_{\psi}\left\|\psi\right\|_2,
    \label{loss:smplx reg}
\end{equation}
where $m_h$, $\beta_b$, $\theta_f$ denote the hand, body, and facial pose latent parameters, $\beta$ refers to body shape parameters, and $\psi$ is the facial expression parameters, as introduced in SMPL-X~\cite{SMPL-X:2019}.
The full loss used to optimize SMPL-X parameters is formally defined as:
\begin{equation}
    \mathcal{L}_{fitting} = \lambda_b\mathcal{L}_{b2d} + \lambda_h\mathcal{L}_{h2d} + \mathcal{L}_{reg},
\end{equation}
where $\mathcal{L}_{h2d}$ is the projection loss for hand joints, and $\mathcal{L}_{b2d}$ is the projection loss for the rest body joints. We set $\lambda_b=1e-3$, $\lambda_h=1e-3$, $\lambda_m=1$, $\lambda_l=1e-3$, $\lambda_f=1e$, $\lambda_\beta=3e-2$, $\lambda_\psi=1e-1$ to balance the loss terms. In order to maintain temporal consistency, we use the optimized SMPL-X parameters from the previous frame as initialization. As demonstrated in Fig.~\ref{fig:smplx}, our fitting results are better aligned with the actual human pose.
\subsection{Skin Weight Transfer}
We animate the body part of the mesh (non-garment region) using Linear Blend Skinning (LBS), employing a robust skin weight transfer algorithm through weight inpainting as introduced by Abdrashitov \etal~\cite{abdrashitov2023robust}. This method effectively transfers skin weights from the SMPL-X~\cite{SMPL-X:2019} mesh (source) to the human mesh (target) in two stages. Initially, we directly transfer skinning weights to target vertices that meet the closest point matching criteria, thoroughly discussed in Section 3.1 of Abdrashitov \etal~\cite{abdrashitov2023robust}. Subsequent to this direct transfer, for vertices that remain without skinning weights, we apply weight inpainting. This step involves extrapolating the missing weights to ensure a seamless and natural skin deformation across the mesh, treated as an optimization problem and described in Section 3.2 of Abdrashitov \etal~\cite{abdrashitov2023robust}. We refer the reader to Abdrashitov \etal~\cite{abdrashitov2023robust} for a detailed explanation, where the processes and their applications are comprehensively described.

\subsection{Garment Physics Parameter Estimation}
We outline the pseudo-code for estimating garment physics parameters in Algorithm \ref{alg:physics} and provide details of the simulator and boundary node specifications below.
\subsubsection{Simulator}  For simulating garment dynamics, we adopt the open-source C-IPC simulator \cite{li2021codimensional}, available at \hyperlink{https://github.com/ipc-sim/Codim-IPC}{https://github.com/ipc-sim/Codim-IPC}. In the task of garment material estimation, we opt for a sequence length of 24 frames, balancing speed with the need for sufficient dynamic information. A longer sequence, while containing more dynamic information, demands prohibitive simulation times, making it impractical for optimization tasks. Conversely, a shorter sequence might fail to capture adequate dynamics, offering insufficient data for material estimation. The garment simulation runs around $0.05-1$ fps on the CPU.
\subsubsection{Boundary Nodes Specification} In our setup, occlusions within the garment mesh—such as areas near the shoulders obscured by hair—require the designation of specific garment nodes as boundary points, also known as Dirichlet boundary conditions. During simulation, the garment is driven by these boundary nodes in conjunction with the underlying body collider. To identify these boundary points, we focus on regions with minimal relative movement between the garment and the body, such as those near the shoulder. The dynamics of these selected boundary points are then calculated using Linear Blend Skinning (LBS), utilizing the nearest corresponding SMPL skinning weights, to obtain $\Verts^{b}_{1:T}$.
\subsubsection{Hyperparameters}
We configure the simulation with a step size of $\delta t = 0.04$ and optimize the garment physics parameters over 100
iterations using the Adam optimizer. For finite difference-based gradient estimation, we set $\delta \rho = 5$, $\delta \stiffs = 0.05$, and $\delta \stiffb = 0.05$, with $\rho$ ranging from 200 to 640, $\stiffb$ and $\stiffs$ each ranging from 0.1 to 8. The scaling for these parameters is detailed in the C-IPC codebase.

\begin{algorithm}
\caption{Garment Physics Parameter Estimation}
\label{alg:physics}
\textbf{Input:} Tracked mesh sequences $\{\mathbf{V}_{1:T}, \mathbf{F}\}$, Garment Boundary condition $\{\mathbf{V}_{1:T}^b\}$, Body Collider $\{\mathbf{V}_{1:T}^C, \mathbf{F}^C\}$,  initial garment density $\rho$, initial garment membrane stiffness $\stiffs$, initial bending stiffness $\stiffb$, step size $\delta t$, density step size $\delta_\rho$, bending stiffness step size $\delta_{\stiffb}$, membrane stiffness step size $\delta_{\stiffs}$, the optimizer $\text{Adam}(\cdot)$, and the cloth simulator  $f(\cdot)$.

\textbf{Output:} $\rho$, $\stiffs$, $\stiffb$

\medskip

\medskip

\SetKwFunction{FMain}{get\_loss}
\SetKwProg{Fn}{Function}{:}{}
\Fn{\FMain{$\rho, \stiffs, \stiffb$}}{

    $\Loss{sim} = 0$
    
    $\dot{\Verts}^g_{0} = 0$

    $\Verts^g_{0} = \Verts_{0}$
    
    \For{$0 \leq t \leq$ T}
    {$\Verts^g_{t+1} = f(\Verts^g_{t}, \dot{\Verts}^g_{t}, \Verts_{t+1}^b, \mathbf{V}^C_{t+1}, \rho, \stiffs, \stiffb, \Delta t)$

    $\Loss{sim} = \Loss{sim} + \textbf{MSE}(\Verts^g_{t+1}, \Verts^g_{t})$
    }
    
    \textbf{return} $\Loss{sim}$
}

\medskip

\For{$0 \leq i \leq$ itermax}{
    \medskip
    
    \medskip
    
    \text{\#Apprximate Gradient using finite difference}

    $\Delta \rho  = \left( \FMain(\rho+\delta_\rho, \stiffs, \stiffb) - \FMain(\rho, \stiffs, \stiffb) \right) / {\delta_\rho}$

    $\Delta \stiffs = \left( \FMain(\rho, \stiffs+\delta_{\stiffs}, \stiffb) - \FMain(\rho, \stiffs, \stiffb) \right) / {\delta_{\stiffs}}.$

    $\Delta \stiffb = \left( \FMain(\rho, \stiffs, \stiffb+\delta_{\stiffb}) - \FMain(\rho, \stiffs, \stiffb) \right) / {\delta_{\stiffb}}.$
    
    \medskip

    \medskip
    
    \text{\#Update the parameter}
    
    $\rho = \text{ Adam}_\rho(\Delta \rho)$
    
    $\stiffs = \text{ Adam}_{\stiffs}(\Delta \stiffs)$
    
    $\stiffb = \text{ Adam}_{\stiffb}(\Delta \stiffb)$
    
    }
\end{algorithm}

\subsection{Appearance Modeling}
\subsubsection{Lighting}
With the ActorHQ dataset \cite{isik2023humanrf}, we assume that lighting conditions can be approximated by constant environmental lighting, represented by a global ambient light parameterized by a global $L_a$. We initialize $L_a$ to have a unit power per unit area per unit steradian across all three color channels.
\subsubsection{Rendering}
We utilize the differentiable path tracer Mitsuba 3\cite{nimier2019mitsuba}, leveraging GPU backends for accelerated rendering. During optimization, we apply 8 samples per pixel (spp) for efficiency. For final rendering, this is increased to 128 spp for enhanced quality. Relighting and additional rendering tasks are conducted in Blender\cite{blender}.
\subsubsection{Optimization Details} 
For inverse rendering, we employ the Adam optimizer, conducting optimization for 2000 iterations for each character. At each gradient step, a timestep and camera view are randomly selected. Supervision is applied within the masked region only for each character. Given the imperfection of the ground-truth masks, we also erode the ground-truth mask by 2 pixels to mitigate potential artifacts. We use $\text{lr}=0.01$ and decay the learning rate by 0.5 every 500 iterations. We clip the optimized texture after each gradient update to ensure legal color values.

\begin{figure}[!t]
    \centering
    \includegraphics[width=0.95\linewidth]{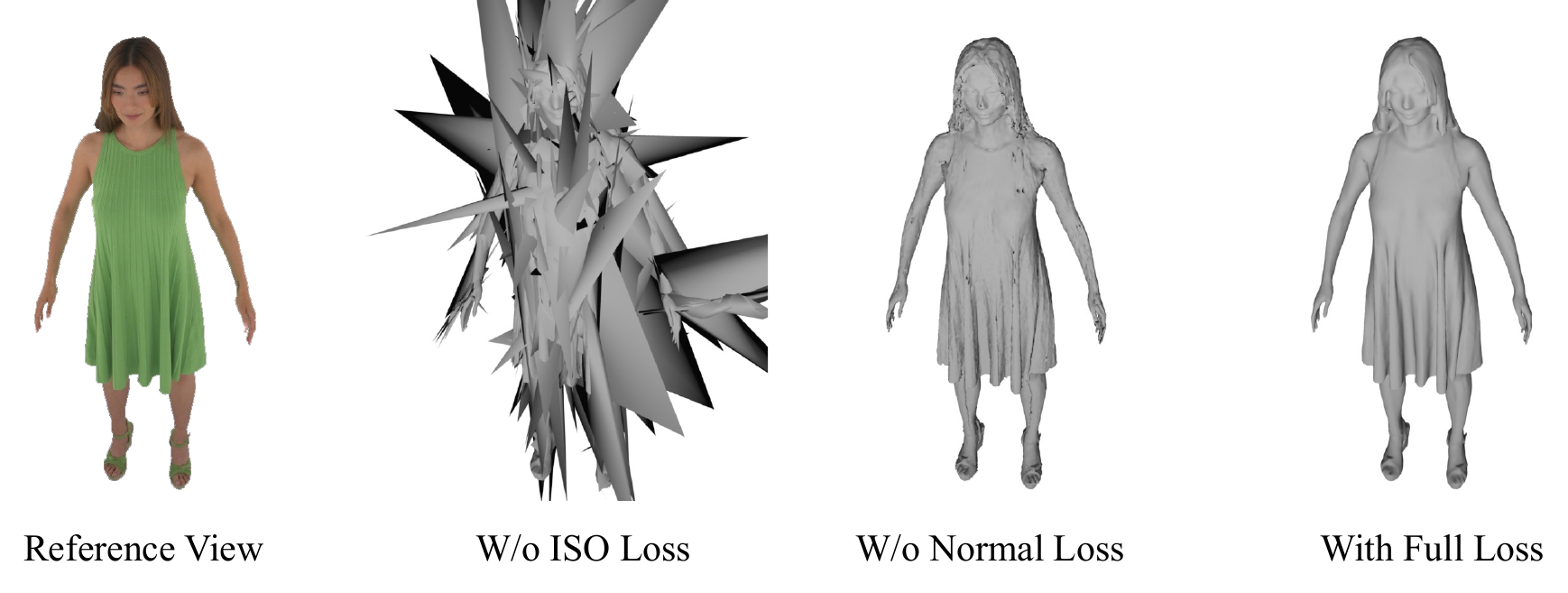}
    \caption{Ablation Study on loss terms used in mesh tracking. Without the ISO loss, the mesh topology gradually becomes chaotic (W/o ISO Loss). Without the normal loss, the mesh surface becomes noisy during optimization (W/o Normal Loss). With the full loss terms, we can obtain a smooth and robust mesh tracking result.}
    \vspace{-0pt}
    \label{fig:ab_tracking}
\end{figure}
\section{Additional Results}
\subsection{Ablation Study On Mesh Tracking}
As introduced in Sec. 3 of the main paper, we propose a robust mesh tracking method to capture accurate geometry correspondences, which is crucial for the subsequent physical parameter estimations. To validate the effectiveness of our approach, we conduct an ablation study focusing on the various loss terms employed during the mesh tracking process.
As demonstrated in Fig.~\ref{fig:ab_tracking}, the ISO loss is important for preserving the local mesh topology. This ensures that the mesh deformation remains realistic by maintaining edge lengths, thereby preventing distortions. Additionally, our introduced normal loss is a key component to generating smooth and robust tracking results. Moreover, we don't observe significant contributions for the regularization losses aside from the ISO loss introduced in Dynamic 3D Gaussians~\cite{luiten2023dynamic} (e.g., rigidity loss and rotation loss), which might not be necessary in our setting.
\subsection{Additional Animation Results}
We show additional animation results on our website: \url{https://qingqing-zhao.github.io/PhysAvatar}. Please refer to it for better visualization. 

\newpage
{\small
\bibliographystyle{splncs04}
\bibliography{egbib}
}
\end{document}


\title{Supplementary Material - PhysAvatar: Learning the Physics of Dressed 3D Avatars from Visual Observations}

\titlerunning{PhysAvatar}

\author{}

\authorrunning{}

\institute{}

\maketitle

%
%
\section{Implementation Details}
\subsection{SMPL-X Fitting}
\begin{figure}[t!]
    \centering
    \includegraphics[width=\linewidth]{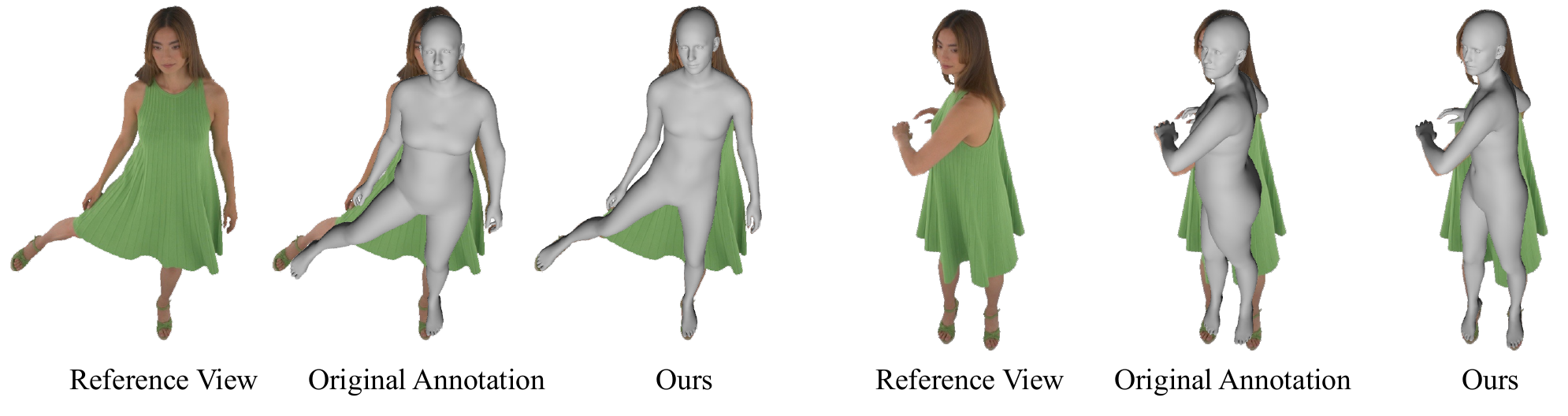}
    \vspace{-10pt}
    \caption{Our SMPL-X~\cite{SMPL-X:2019} fitting results show better accuracy compared to the original annotations from Actor-HQ~\cite{isik2023humanrf} dataset.}
    \label{fig:smplx}
\end{figure}

We observe that the human pose estimations provided by Actor-HQ~\cite{isik2023humanrf} are noisy, and failure cases exist in some sequences with challenging poses. Since our method relies on a body collider that provides accurate contact information for garment simulation, an SMPL-X mesh that aligns well with the human pose and shape is necessary. To this end, we fit the SMPL-X parameters from multi-view videos using an algorithm~\cite{zhang2021lightweight, zheng2021deepmulticap} simplified for our setting. Specifically, given image $\Image_{i,t}$ captured from camera $i$ at $t$ timestamp, we detect 2D human keypoints $\mathbf{x}_{i, t}$ using state-of-the-art human keypoints detector DWPose~\cite{mmpose2020, yang2023effective} and calculate the projection loss as follows:
\begin{equation}
    \mathcal{L}_{2d} = \left\| \mathbf{x}_{i, t} - \hat{\mathbf{x}}_{i, t} \right\|_2,
    \label{loss:projection}
\end{equation}
where $\hat{\mathbf{x}}_{i, t}$ is the projection of SMPL-X 3D joints under camera $i$. We also employ a regularization term as:
\begin{equation}
    \mathcal{L}_{reg} = \lambda_m\left\|m_h\right\|_2 + \lambda_{l}\left\|\theta_b\right\|_2 + \lambda_{f}\left\|\theta_f\right\|_2 + 
    \lambda_{\beta}\left\|\beta\right\|_2 + \lambda_{\psi}\left\|\psi\right\|_2,
    \label{loss:smplx reg}
\end{equation}
where $m_h$, $\beta_b$, $\theta_f$ denote the hand, body, and facial pose latent parameters, $\beta$ refers to body shape parameters, and $\psi$ is the facial expression parameters, as introduced in SMPL-X~\cite{SMPL-X:2019}.
The full loss used to optimize SMPL-X parameters is formally defined as:
\begin{equation}
    \mathcal{L}_{fitting} = \lambda_b\mathcal{L}_{b2d} + \lambda_h\mathcal{L}_{h2d} + \mathcal{L}_{reg},
\end{equation}
where $\mathcal{L}_{h2d}$ is the projection loss for hand joints, and $\mathcal{L}_{b2d}$ is the projection loss for the rest body joints. We set $\lambda_b=1e-3$, $\lambda_h=1e-3$, $\lambda_m=1$, $\lambda_l=1e-3$, $\lambda_f=1e$, $\lambda_\beta=3e-2$, $\lambda_\psi=1e-1$ to balance the loss terms. In order to maintain temporal consistency, we use the optimized SMPL-X parameters from the previous frame as initialization. As demonstrated in Fig.~\ref{fig:smplx}, our fitting results are better aligned with the actual human pose.
\subsection{Skin Weight Transfer}
We animate the body part of the mesh (non-garment region) using Linear Blend Skinning (LBS), employing a robust skin weight transfer algorithm through weight inpainting as introduced by Abdrashitov \etal~\cite{abdrashitov2023robust}. This method effectively transfers skin weights from the SMPL-X~\cite{SMPL-X:2019} mesh (source) to the human mesh (target) in two stages. Initially, we directly transfer skinning weights to target vertices that meet the closest point matching criteria, thoroughly discussed in Section 3.1 of Abdrashitov \etal~\cite{abdrashitov2023robust}. Subsequent to this direct transfer, for vertices that remain without skinning weights, we apply weight inpainting. This step involves extrapolating the missing weights to ensure a seamless and natural skin deformation across the mesh, treated as an optimization problem and described in Section 3.2 of Abdrashitov \etal~\cite{abdrashitov2023robust}. We refer the reader to Abdrashitov \etal~\cite{abdrashitov2023robust} for a detailed explanation, where the processes and their applications are comprehensively described.

\subsection{Garment Physics Parameter Estimation}
We outline the pseudo-code for estimating garment physics parameters in Algorithm \ref{alg:physics} and provide details of the simulator and boundary node specifications below.
\subsubsection{Simulator}  For simulating garment dynamics, we adopt the open-source C-IPC simulator \cite{li2021codimensional}, available at \hyperlink{https://github.com/ipc-sim/Codim-IPC}{https://github.com/ipc-sim/Codim-IPC}. In the task of garment material estimation, we opt for a sequence length of 24 frames, balancing speed with the need for sufficient dynamic information. A longer sequence, while containing more dynamic information, demands prohibitive simulation times, making it impractical for optimization tasks. Conversely, a shorter sequence might fail to capture adequate dynamics, offering insufficient data for material estimation. The garment simulation runs around $0.05-1$ fps on the CPU.
\subsubsection{Boundary Nodes Specification} In our setup, occlusions within the garment mesh—such as areas near the shoulders obscured by hair—require the designation of specific garment nodes as boundary points, also known as Dirichlet boundary conditions. During simulation, the garment is driven by these boundary nodes in conjunction with the underlying body collider. To identify these boundary points, we focus on regions with minimal relative movement between the garment and the body, such as those near the shoulder. The dynamics of these selected boundary points are then calculated using Linear Blend Skinning (LBS), utilizing the nearest corresponding SMPL skinning weights, to obtain $\Verts^{b}_{1:T}$.
\subsubsection{Hyperparameters}
We configure the simulation with a step size of $\delta t = 0.04$ and optimize the garment physics parameters over 100
iterations using the Adam optimizer. For finite difference-based gradient estimation, we set $\delta \rho = 5$, $\delta \stiffs = 0.05$, and $\delta \stiffb = 0.05$, with $\rho$ ranging from 200 to 640, $\stiffb$ and $\stiffs$ each ranging from 0.1 to 8. The scaling for these parameters is detailed in the C-IPC codebase.
\begin{algorithm}
\caption{Garment Physics Parameter Estimation}
\label{alg:physics}
\textbf{Input:} Tracked mesh sequences $\{\mathbf{V}_{1:T}, \mathbf{F}\}$, Garment Boundary condition $\{\mathbf{V}_{1:T}^b\}$, Body Collider $\{\mathbf{V}_{1:T}^C, \mathbf{F}^C\}$,  initial garment density $\rho$, initial garment membrane stiffness $\stiffs$, initial bending stiffness $\stiffb$, step size $\delta t$, density step size $\delta_\rho$, bending stiffness step size $\delta_{\stiffb}$, membrane stiffness step size $\delta_{\stiffs}$, the optimizer $\text{Adam}(\cdot)$, and the cloth simulator  $f(\cdot)$.

\textbf{Output:} $\rho$, $\stiffs$, $\stiffb$

\medskip

\medskip

\SetKwFunction{FMain}{get\_loss}
\SetKwProg{Fn}{Function}{:}{}
\Fn{\FMain{$\rho, \stiffs, \stiffb$}}{

    $\Loss{sim} = 0$
    
    $\dot{\Verts}^g_{0} = 0$

    $\Verts^g_{0} = \Verts_{0}$
    
    \For{$0 \leq t \leq$ T}
    {$\Verts^g_{t+1} = f(\Verts^g_{t}, \dot{\Verts}^g_{t}, \Verts_{t+1}^b, \mathbf{V}^C_{t+1}, \rho, \stiffs, \stiffb, \Delta t)$

    $\Loss{sim} = \Loss{sim} + \textbf{MSE}(\Verts^g_{t+1}, \Verts^g_{t})$
    }
    
    \textbf{return} $\Loss{sim}$
}

\medskip

\For{$0 \leq i \leq$ itermax}{
    \medskip
    
    \medskip
    
    \text{\#Apprximate Gradient using finite difference}

    $\Delta \rho  = \left( \FMain(\rho+\delta_\rho, \stiffs, \stiffb) - \FMain(\rho, \stiffs, \stiffb) \right) / {\delta_\rho}$

    $\Delta \stiffs = \left( \FMain(\rho, \stiffs+\delta_{\stiffs}, \stiffb) - \FMain(\rho, \stiffs, \stiffb) \right) / {\delta_{\stiffs}}.$

    $\Delta \stiffb = \left( \FMain(\rho, \stiffs, \stiffb+\delta_{\stiffb}) - \FMain(\rho, \stiffs, \stiffb) \right) / {\delta_{\stiffb}}.$
    
    \medskip

    \medskip
    
    \text{\#Update the parameter}
    
    $\rho = \text{ Adam}_\rho(\Delta \rho)$
    
    $\stiffs = \text{ Adam}_{\stiffs}(\Delta \stiffs)$
    
    $\stiffb = \text{ Adam}_{\stiffb}(\Delta \stiffb)$
    
    }
\end{algorithm}

\subsection{Appearance Modeling}
\subsubsection{Lighting}
With the ActorHQ dataset \cite{isik2023humanrf}, we assume that lighting conditions can be approximated by constant environmental lighting, represented by a global ambient light parameterized by a global $L_a$. We initialize $L_a$ to have a unit power per unit area per unit steradian across all three color channels.
\subsubsection{Rendering}
We utilize the differentiable path tracer Mitsuba 3\cite{nimier2019mitsuba}, leveraging GPU backends for accelerated rendering. During optimization, we apply 8 samples per pixel (spp) for efficiency. For final rendering, this is increased to 128 spp for enhanced quality. Relighting and additional rendering tasks are conducted in Blender\cite{blender}.
\subsubsection{Optimization Details} 
For inverse rendering, we employ the Adam optimizer, conducting optimization for 2000 iterations for each character. At each gradient step, a timestep and camera view are randomly selected. Supervision is applied within the masked region only for each character. Given the imperfection of the ground-truth masks, we also erode the ground-truth mask by 2 pixels to mitigate potential artifacts. We use $\text{lr}=0.01$ and decay the learning rate by 0.5 every 500 iterations. We clip the optimized texture after each gradient update to ensure legal color values.





    
    


\begin{figure}[!t]
    \centering
    \includegraphics[width=0.95\linewidth]{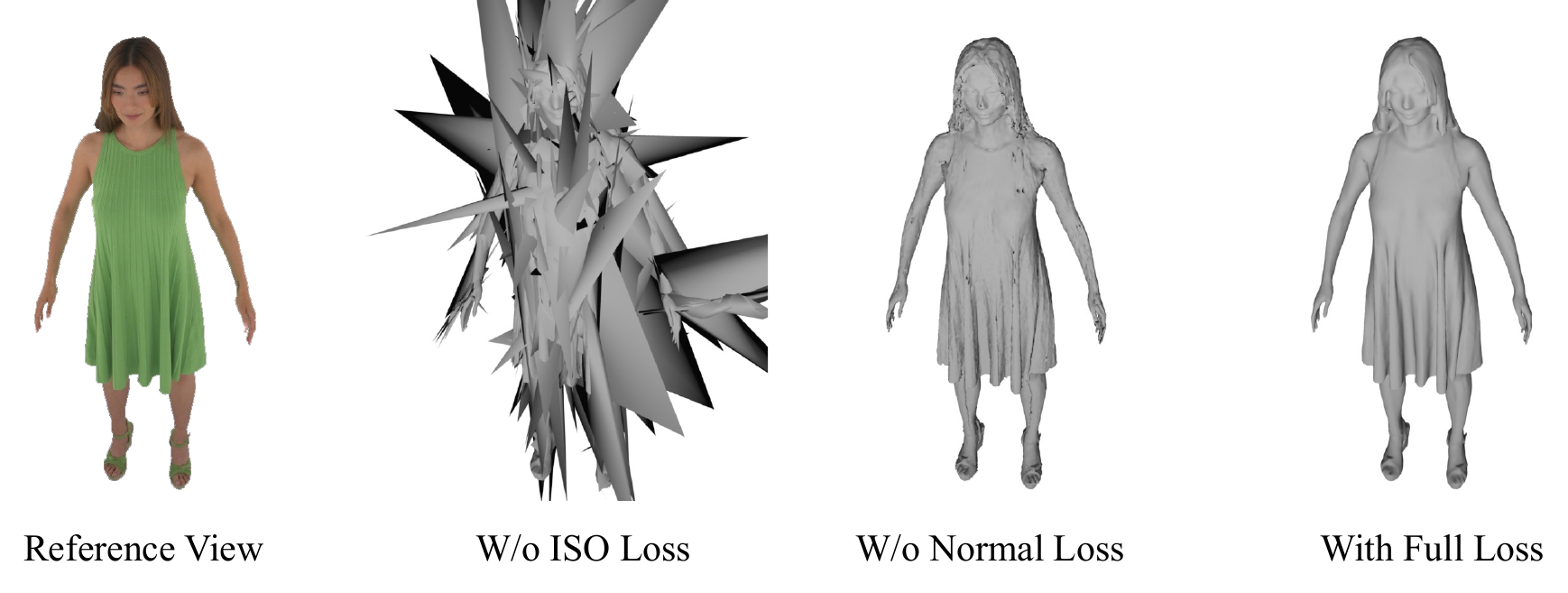}
    \caption{Ablation Study on loss terms used in mesh tracking. Without the ISO loss, the mesh topology gradually becomes chaotic (W/o ISO Loss). Without the normal loss, the mesh surface becomes noisy during optimization (W/o Normal Loss). With the full loss terms, we can obtain a smooth and robust mesh tracking result.}
    \vspace{-0pt}
    \label{fig:ab_tracking}
\end{figure}
\section{Additional Results}
\subsection{Ablation Study On Mesh Tracking}
As introduced in Sec. 3 of the main paper, we propose a robust mesh tracking method to capture accurate geometry correspondences, which is crucial for the subsequent physical parameter estimations. To validate the effectiveness of our approach, we conduct an ablation study focusing on the various loss terms employed during the mesh tracking process.
As demonstrated in Fig.~\ref{fig:ab_tracking}, the ISO loss is important for preserving the local mesh topology. This ensures that the mesh deformation remains realistic by maintaining edge lengths, thereby preventing distortions. Additionally, our introduced normal loss is a key component to generating smooth and robust tracking results. Moreover, we don't observe significant contributions for the regularization losses aside from the ISO loss introduced in Dynamic 3D Gaussians~\cite{luiten2023dynamic} (e.g., rigidity loss and rotation loss), which might not be necessary in our setting.
\subsection{Additional Animation Results}
We show additional animation results in our supplementary video. Please refer to the video for better visualization. 
{\small
\bibliographystyle{splncs04}
\bibliography{egbib}
}